\documentstyle[twoside,fleqn,espcrc2,epsfig]{article}

\newcommand{\np}{Nucl.Phys.\ }
\newcommand{\pl}{Phys.Lett.\ }
\newcommand{\pr}{Phys.Rev.\ }
\newcommand{\asgen}{\alpha_s}
\newcommand{\Gev}{{\rm GeV}}

\newcommand{\AmS}{{\protect\the\textfont2
  A\kern-.1667em\lower.5ex\hbox{M}\kern-.125emS}}

\def\np#1#2#3{Nucl.\ Phys.\ B#1 (19#3) #2}
\def\pl#1#2#3{Phys.\ Lett.\ #1B (19#3) #2}
\def\pr#1#2#3{Phys.\ Rev.\ D #1 (19#3) #2}

\title{Search for $\frac{\Lambda^2}{p^2}$ corrections to the
       QCD running coupling}

\author{G. Burgio\address{Dipartimento di Fisica, Universit\`a di Parma
        and INFN, Gruppo Collegato di Parma, Italy},
	F. Di Renzo\address{Department of Mathematical Sciences, 
	University of Liverpool, Liverpool L69 3BX, U.K. (UKQCD Collaboration)}\thanks{presented by F. Di Renzo}, 
	C. Parrinello$^{\rm b}$ 
	and 
	C. Pittori\address{Institut de Physique Nucl\'eaire Th\'eorique,
	Universit\'e de Li\`ege au Sart Tilmann, B-4000 Li\`ege, Belgique}}
       
\begin{document}

\begin{abstract}
We investigate the occurrence of power terms $\frac{\Lambda^2}{p^2}$ 
in the running  QCD coupling by analysing non-perturbative 
measurements of $\asgen(p)$ at quite low momenta  
obtained from the lattice three-gluon vertex. 
Our study provides some evidence for such a contribution. 
The phenomenological implications of such a presence are reviewed. 
\end{abstract}

\maketitle

\section{Introduction and motivations}
The Operator Product Expansion (OPE) is the standard way to parametrise 
non-perturbative QCD effects in terms of power corrections, the powers 
involved beeing uniquely fixed by the operator content of the theory. 
Anyway, due to the asymptotic nature of 
QCD perturbative expansions, i.e. due to renormalons ambiguities, power 
corrections are reshuffled between operators and coefficient functions 
\cite{ren}. 
It has recently been conjectured \cite{CeccoPeppe} that power corrections 
which are not 
{\it a priori} expected from OPE may appear in the expansion of 
physical observables, via much the same analysis as for renormalons, if 
one allows the presence of (UV-subleading) power corrections to $\asgen(p)$. 
The lattice community knows that perturbative logarithms are of course not 
the only contribution to the running coupling. Aim of the present work is 
to search for non-perturbative contributions to the running coupling in the 
form of power corrections, with the given power ($\frac{\Lambda^2}{p^2}$). 
Before briefly reviewing some theoretical arguments in support of this 
{\it theoretical prejudice}, a couple of considerations are in order. 
While power corrections to $\asgen (p)$ arise naturally in many physical 
schemes \cite{pino,maclep} and their 
occurrence cannot be excluded {\em a priori} in any scheme, 
their non-perturbative nature makes it 
very hard to assess their scheme dependence, which is 
only very weakly constrained by the general properties of the theory.

\subsection{A lesson from the Gluon Condensate}
A stage on which all the considerations above apply is provided by the 
efforts towards a lattice determination of the Gluon Condensate, which is 
given in terms of Wilson loops $W$ \cite{CeccoPeppe}. The OPE for W is to 
be written down as
\begin{equation}
W = W_0 + \frac{\Lambda^4 W_4}{Q^4} + \ldots
\end{equation}
$W_0$ beeing the contribution related to the identity operator and the second 
term beeing the ``genuine'' (dim=$4$) condensate ($Q$ is the scale; on the 
lattice $Q \sim 1/a$). Perturbative contributions are present only in $W_0$, 
so that a standard procedure to extract the condensate was to subtract from 
MonteCarlo measurements of Wilson loops their perturbative expansions. From 
general considerations the expected form of $W_0$ is 
\begin{equation}
\label{integrale}
W_0 = \int_{\rho \Lambda^2}^{Q^2} \frac{dp^2}{p^2} (\frac{p^2}{Q^2})^2 
\asgen(\frac{p^2}{\Lambda^2})\; .
\end{equation} 
A renormalon analysis of this formula readily shows that a 
$O(\Lambda^4/Q^4)$ ambiguity is present in $W_0$, thus preventing any 
unambiguous result from the above procedure. The situation is anyway 
even more intricate: actually, after having 
resummed $W_0$ (within the mentioned ambiguity), performing the subtraction 
leaves one with something that scales like $O(\Lambda^2/Q^2)$ 
\cite{lepmac}. The puzzle 
can be sorted out if one supposes in (\ref{integrale}) contributions of the 
form $\asgen(p) \sim \frac{\Lambda^2}{p^2}$. 

\subsection{Static quarks potential and confinement}
A nice argument to support the presence of $\frac{\Lambda^2}{p^2}$ 
contributions to $\asgen(p)$ comes from considerations involving confinement. 
Consider the interaction 
of two heavy quarks in the static limit \cite{Zak}. Within the Born 
approximation one can obtain the static potential from 
\begin{equation}
\label{HQP}
V(r) \, \propto \, \int d^3k \, \alpha_s (|\vec{k}|^2) \,  
\frac{\exp^{i \vec{k} \cdot \vec{r}}}{|\vec{k}|^2}.
\end{equation}
The above formula has been written down in the renormalon-style: plugging 
in a constant $\asgen$ yields the Coulomb potential ($V(r) \approx 1/r$), 
while the leading-logs expression for $\asgen$ generates
various power corrections to the potential; however the string 
tension contribution $V(r) \sim K r$ is missing. Such a contribution is 
obtained by plugging in a power correction to $\asgen$ of the form we 
are looking for. Notice that the above considerations are in a sense 
not new in the context of non-perturbative 
contributions to the running coupling \cite{Chris}. 
Consider the ``force" definition of the running coupling:
\begin{equation}
\alpha_{q\bar{q}}(Q) = \frac{3}{4} r^2 \frac{dV}{dr} \;\;\;\;\;
(Q = \frac{1}{r}),
\end{equation}
where again $V(r)$ is the static interquark potential.
By keeping into account the string tension contribution to $V(r)$, one 
obtains a $1/Q^2$ contribution, whose order of magnitude is given 
by the string tension itself. While this term was mainly considered 
as a sort of ambiguity, resulting in an indetermination in the value 
of $\alpha(Q)$ at a given scale, it can be interpreted as a clue 
for the existence of a 
$\frac{\Lambda^2}{p^2}$ contribution, providing also an estimate 
for the expected order of magnitude of it, at least in one (physically sound) 
scheme. 


\section{Fits to lattice data}
The lattice data for $\asgen(p)$ that we used for our analysis were obtained 
by evaluating two- and three-point off-shell Green's functions of the gluon 
field in the Landau gauge and imposing non-perturbative renormalisation 
conditions on them, for different values of the external momenta \cite{cla}. 
Such a definition of the coupling corresponds to a 
momentum-subtraction renormalisation scheme in continuum QCD \cite{mom}. 
The data available at the moment are quite noisy and lattice artifacts are 
still to be fully assessed (they seem under control from the analysis 
in \cite{cla}). Given the particular ansatz to which we want to fit the 
data (perturbative expressions of given order plus a power correction), a 
peculiar momentum interval has to be singled out. On one hand, 
the momentum range should start well above the location of the perturbative 
Landau pole, but it should nonetheless include low scales where 
power corrections may still be sizeable. On the other hand, the 
requirement of keeping the effects of the finite lattice spacing under 
control in the numerical data for $\asgen$ induces a natural UV cutoff. 


\subsection{Two-loop analysis}
At two-loop level we fit data to the formula
\begin{equation}
\alpha_s(p) = \frac{1}{b_0 \, L_2} - \frac{b_1}{b_0} 
\frac{\log(L_2)}{(b_0 \, L_2)^2} + c_{2l} \, \frac{\Lambda_{2l}^2}{p^2}
\end{equation}
where $L_2 = \log(p^2/\Lambda_{2l}^2)$. Notice that since the value of 
$\Lambda$ is expected to carry a sizeable dependence on the order of the 
perturbative calculation, we append a subscript. We were able to check that 
out of power corrections of the form $(\frac{\Lambda_{2l}^2}{p^2})^z$ the 
value $z \approx 1$ is indeed the best choice, which thing we interpret 
as a confirmation of our theoretical prejudice. Our best fit singles out 
the values ($\Lambda_{2l}=0.84(1)$,$c_{2l}=0.31(3)$), with a $\chi^2_{dof} 
\leq 1.8$. The momentum interval which is best described is 
$p \sim 2 - 3$ \Gev, fully consistent with the requirements we have already 
mentioned.

\subsection{Three-loop analysis}
Given the possible interplay between a description in terms of power 
corrections and our ignorance about higher orders, a three-loop analysis is 
compelling. A major obstacle for such an analysis is actually the fact 
that the first non-universal $\beta$-function coefficient $b_2$ is not 
known for our scheme. 
Therefore we also fitted data to the formula 
\begin{eqnarray}
\protect\label{3lp}
\alpha_s(p) & = & \frac{1}{b_0 \, L_3} \, - \, 
\frac{b_1}{b_0} 
\frac{\log(L_3)}{(b_0 \, L_3)^2} + \, \frac{1}{(b_0 \, L_3)^3} \, \nonumber \\
& &  \left( \frac{b_2^{eff}}{b_0} + \frac{b_1^2}{b_0^2} 
(\log^2(L_3) - \log(L_3) + 1 ) \right) \nonumber \\
& &  + \, c_{3l} \, \frac{\Lambda_{3l}^2}{p^2},
\end{eqnarray}
where $L_3 = \log(p^2/\Lambda_{3l}^2)$ and $b_2^{eff}$ is an effective 
$\beta$-function coefficient to be determined from the fit. 
In order to gain insight, we started by putting $c_{3l}=0$, obtaining 
the values ($\Lambda_{3l} = 0.72(1)$,$b_2^{eff} = 1.3(1)$), with 
$\chi^2_{dof} \approx 1.8$, in a momentum range $p \sim 2 - 3$ \Gev 
(dashed line in the figure). 
The value of $b_2^{eff}$ is expected to be a reliable 
estimate of $b_2$, as confirmed by 
simple arguments concerning the convergence properties of the expansion 
of our coupling in powers of other couplings. What is more important, 
the value of 
$\Lambda_{3l}$ agrees with other determinations via pertubative matching 
\cite{alphaC}. By keeping into account also $c_{3l}$, moving around the 
same range for $\Lambda_{3l}$ one obtains the values 
($\Lambda_{3l} = 0.72(1)$, $b_2^{eff} = 1.0(1)$, $c_{3l} = 0.41(2)$), 
yielding a $\chi^2_{dof} \approx 1.8$ in a momentum range 
$p \sim 1.8 - 3$ \Gev (solid line in the figure). 
While $b_2^{eff}$ still makes sense, it emerges 
that $c_{2l} \Lambda_{2l}^2 = 0.22(2) \, \Gev^2 \sim c_{3l} \Lambda_{3l}^2 = 0.21(2)
 \, \Gev^2$, which is just the order of the standard estimate for 
the string tension. 

\begin{figure}[htb]
\begin{center}
\mbox{{\epsfig{figure=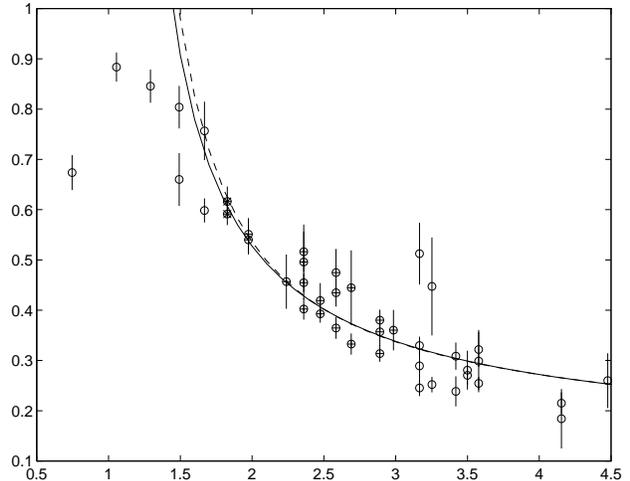,height=6.5 cm}}}
\end{center}
\caption{Three-loop fits to lattice data for the coupling, with and without 
power corrections.}
\label{fig:3lp}
\end{figure}

\section{Conclusions and perspectives}
Some preliminary evidence in support of a $\frac{\Lambda^2}{p^2}$ 
contribution to $\asgen(p)$ has been found, which appears at this 
stage disentangled from possible perturbative ambiguities: perturbative and 
non-perturbative (power) contributions do not mix in our formulae  
when upgrading from a two-loop to a three-loop description. 
While our results need to be further tested, still 
they provide evidence in support of the 
conjecture that ``anomalous" (i.e., not accounted for by OPE) 
power corrections into current correlation functions and physical observables 
may be generated by power corrections to $\asgen(p)$.

\end{document}